\documentstyle[aps]{revtex}
\begin{document}
\baselineskip=1.5\normalbaselineskip

\begin{center}
{\Large {\bf Quantum anti-Zeno effect }}\\[\baselineskip] {B. Kaulakys$^{*}$
and V. Gontis$^{\dagger }$}\\[\baselineskip] {Institute of Theoretical
Physics and Astronomy,} \\ {A. Go\v stauto 12, 2600 Vilnius, Lithuania}

\vspace{2\baselineskip}

\parbox{5in}{\small
Prevention of a quantum system's  time evolution by repetitive, frequent
measurements of the system's state has been called the quantum Zeno effect
(or paradox). Here we investigate theoretically and numerically the effect
of repeated measurements on the quantum dynamics of the multilevel
systems that exhibit the quantum localization of the classical chaos.
The analysis is based on the wave function and Schr\"odinger equation,
without introduction of the density matrix. We show how the quantum Zeno
effect in simple few-level systems can be recovered and understood by
formal modeling the measurement effect on the dynamics by randomizing the
phases of the measured states. Further the similar analysis is extended to
investigate of the dynamics of multilevel systems driven by an intense
external force and affected by frequent measurement. We show that frequent
measurements of such quantum systems results in the delocalization of the
quantum suppression of the classical chaos. This result is the opposite of
the quantum Zeno effect. The phenomenon of delocalization of the quantum
suppression and restoration of the classical-like time evolution of these
quasiclassical systems, owing to repetitive frequent measurements, can
therefore be called the {\it 'quantum anti-Zeno effect'}.
From this analysis we furthermore conclude that frequently or continuously
observable quasiclassical systems evolve basically in a classical manner.}
\end{center}

PACS number(s): 03.65Bz, 05.45.+b, 42.50.Lc

\vskip\baselineskip
\hspace{4cm}

---------------------------------------

$^{*}$ Electronic address: kaulakys@itpa.lt

$^{\dagger }$ Electronic address: gontis@itpa.lt

\newpage

\begin{center}
{\bf I. INTRODUCTION }
\end{center}

Dynamics of a quantum system, which it is not being observed, can be
described by the Schr\"odinger equation. In the von Neumann axiomatics of
quantum mechanics it is postulated that any measurement gives rise to an
abrupt change of the state of the system under consideration and projects it
onto an eigenstate of the measured observable. The measurement process
follows irreversible dynamics, e.g. due to coupling with the multitude of
vacuum modes if spontaneous radiation is registered, and causes the
disappearance of coherence of the system's state: to the decay of the
off-diagonal matrix elements of the density matrix or randomization of the
phases of the wave function's amplitudes.

It is known that a quantum system undergoes relatively slow (Gaussian,
quadratic or cosine type but not exponential) evolution at an early period
after preparation or measurement [1]. Therefore, the repetitive frequent
observation of the quantum system can inhibit the decay of unstable [2]
system and suppress dynamics of the driven by an external field [3, 4]
system. This phenomenon, namely the inhibition or even prevention of the
time evolution of the system from an eigenstate of observable into a
superposition of eigenstates by repeated frequent measurement, is called the
quantum Zeno effect (paradox) or the quantum watched pot [2-5]. Usually
derivation and investigation of the quantum Zeno effect is based on the von
Neumann's postulate of projection or reduction of the wave-packet in the
measurement process. However, the outcome of the variation of the quantum
Zeno effect in a three-level system, originally proposed by Cook [3] and
experimentally realized by Itano {\it et al.} [4] has been explained by
Frerichs and Schenzle [6] on the basis of the standard three-level Bloch
equations for the density matrix in the rotating-wave approximation with the
spontaneous relaxation. Thus, the quantum Zeno effect can be derived either
from the {\it ad hoc} collapse hypothesis [2-4] or formulated in terms of
irreversible quantum dynamics without additional assumptions, i.e. as the
dynamical quantum Zeno effect [6,7]. Moreover, the postulate of the
'collapse of the wave function' models the actual measurement process only
roughly [6].

Aharonov and Vardi [8] showed that frequent measurements can not only stop
the quantum dynamics but it also can induce time evolution of the observable
system. They used the von Neumann projection postulate and predicted an
evolution of the system along a presumed trajectory due to a sequence of
measurements performed on states that slightly change from measurement to
measurement. Altenm\"uller and Schenzle [9] have demonstrated that such a
phenomenon can be realized replacing the collapse hypothesis by an
irreversible physical interaction.

It should be noted that most of the systems analyzed in the papers mentioned
above consist only of the few (usually two or three) quantum states and are
purely quantum. Consequently, it is of interest to investigate the influence
of the repeated frequent measurement on the evolution of the multilevel
quasiclassical systems, the classical counterparts of which exhibit chaos.
It has been established [10-12] that chaotic dynamics of such systems, e.g.
dynamics of nonlinear systems strongly driven by a periodic external field,
is suppressed by the quantum interference effect and it gives rise to the
quantum localization of the classical dynamics in the energy space of the
system. Thus, the quantum localization phenomenon strongly limits the
quantum motion. As it was stated above, the repeated frequent measurement or
continuous observation of the quantum system can inhibit its dynamics as
well. Therefore, it is natural to expect that frequent measurement of the
suppressed system will result in the additional freezing of the system's
state.

In connection with this question we should refer to the papers where the
influence of small external noise, environment and measurement induced
effects on the quantum chaos is analyzed (see [13-20] and references
therein). The general conclusion of such investigations is that noise,
interaction with the environment and measurement induce the decoherence,
irreversibility and delocalization. However, the direct link between
measurements of the suppressed chaotic systems and the quantum Zeno effect,
to the best of our knowledge, have not yet been analyzed. We can only refer
to papers [21] where some preliminary relation between the quantum Zeno
effect and the influence of repeated measurement on the dynamics of the
localized quantum system is presented.

The purpose of this paper is to investigate theoretically and numerically
the influence of dense measurement on the evolution of the multilevel
quasiclassical systems.

The analysis of the measurement effect on the dynamics of the quantum
systems is usually performed with the aid the density matrix formalism.
However, the investigation of the quantum dynamics of the multilevel systems
affected by repeated measurements is very difficult analytically and much
time consuming in numerical calculations. The analysis based on the wave
function and Schr\"odinger equation is considerably easier tractable and
more evident. So, first we will show how the quantum Zeno effect in a
few-level-system can be described in terms of the wave function and
Schr\"odinger equation without introducing of the density matrix and how the
measurements can be incorporated into the equations of motion.

Further we will use the same method for the analysis of the dynamics of the
multilevel system affected by repeated frequent measurement. We reveal that
repetitive measurement of the multilevel systems with quantum suppression of
classical chaos results in the delocalization of the states superposition
and restoration of the chaotic dynamics. Since this effect is reverse to the
quantum Zeno effect we call this phenomenon the {\it 'quantum anti-Zeno
effect'}.

\begin{center}
{\bf II. DYNAMICS\ OF\ TWO-LEVEL\ SYSTEM }
\end{center}

Let's consider the simplest quantum dynamical process and the influence of
frequent measurements on the outcome of the dynamics. Time evolution of the
amplitudes $a_1(t)$ and $a_2\left( t\right) $ of the two-state wave function
$$
\Psi =a_1(t)\Psi _1+a_2\left( t\right) \Psi _2\eqno{(2.1)}
$$
of the system in the resonance field (in the rotating wave approximation) or
of the spin-half system in a constant magnetic field can be represented as
$$
a_1(t)=a_1(0)\cos \frac 12\Omega t+ia_2(0)\sin \frac 12\Omega t
$$
$$
a_2(t)=ia_1(0)\sin \frac 12\Omega t+a_2(0)\cos \frac 12\Omega t,\eqno{(2.2)}
$$
where $\Omega $ is the Rabi frequency. We introduce a matrix ${\bf A}$
representing time evolution during the time interval $\tau $ (between time
moments $t=k\tau $ and $t=(k+1)\tau $ with integer $k$) and rewrite Eq.
(2.2) in the mapping form
$$
\left( \matrix{a_1(k+1)\cr a_2(k+1)}\right) ={\bf A}\left(
\matrix{a_1(k)\cr a_2(k)}\right) \eqno(2.3)
$$
where the evolution matrix ${\bf A}$ is given by
$$
{\bf A}=\left(
\matrix{\cos\varphi& i\sin\varphi\cr
i\sin\varphi & \cos\varphi}\right) ,~~~\varphi ={\frac 12}\Omega \tau .
\eqno(2.4)
$$
Evidently, the evolution of the amplitudes from time $t=0$ to $t=T=n\tau $
can be expressed as
$$
\left( \matrix{a_1(n)\cr a_2(n)}\right) ={\bf A}^n\left(
\matrix{a_1(0)\cr
a_2(0)}\right) .\eqno(2.5)
$$
One can calculate matrix ${\bf A}^n$ by the method of diagonalization of the
matrix ${\bf A.}$ The result naturally is
$$
{\bf A}^n=\left(
\matrix{\cos n\varphi & i\sin n\varphi\cr
i\sin n\varphi & \cos n\varphi}\right) .\eqno(2.6)
$$
Note that $n\varphi =\frac 12\Omega T.$

Equations (2.2)--(2.6) represent time evolution of the system without the
intermediate measurements in the time interval $0\div T$. If at $t=0$ the
system was in the state $\Psi _1$, i.e. $a_1(0)=1$ and $a_2(0)=0$, and if $
\Omega T=\pi $ then at the time moment $t=T$ we would certainly find the
system in the state $\Psi _2$, i.e. it would be $\left| a_1(T)\right| ^2=0$
and $\left| a_2(T)\right| ^2=1$, a certain (with the probability $1$)
transition between the states. Note, that such quantum dynamics without the
intermediate measurements is regular and coherent for all time until the
final measurement.

Let's consider now the dynamics of the system with the intermediate
measurements every time interval $\tau $. Measurement of the system's state
in the time moment $t=k\tau $ projects the system into the state $\Psi _1$
with the probability $p_1(k)=\mid a_1(k)\mid ^2$ or into the state $\Psi _2$
with the probability $p_2(k)=\mid a_2(k)\mid ^2$. After the measurement we
know the probabilities $p_1(k)$ and $p_2(k)$ but we have no information
about the phases $\alpha _1(k)$ and $\alpha _2(k)$ of the amplitudes
$$
a_1(k)=\left| a_1(k)\right| e^{i\alpha _1(k)},~~~a_2(k)=\left| a_2(k)\right|
e^{i\alpha _2(k)},\eqno{(2.7)}
$$
i.e. the phases $\alpha _1(k)$ and $\alpha _2(k)$ after every act of the
measurement are random. Randomization of the phases after the measurement
act can also be confirmed by the analysis of the definite measurement
process, e.g. in the ${\bf V}$-shape tree-level configuration with the
spontaneous transition to the ground state [3-7]. Every measurement of the
system's state results in the mutually uncorrelated phases $\alpha _1(k)$
and $\alpha _2(k)$. After the full measurement of the system's state these
phases are uncorrelated with the phases of the amplitudes before the
measurement too. That is why, according to the measurement postulate the
outcome of the measurement does not depend on the phases of the amplitudes
in the expansion of the wave function through the eigenfunctions of the
measured observation. This will result in the absence of the influence of
the interference terms in the expressions derived below on the transition
probabilities between the eigenstates.

Now we derive equations for the transition probabilities between the states
in the case of evolution with intermediate measurements. From equations
(2.3) and (2.4) we have
$$
\mid a_1(k+1)\mid ^2=\mid a_1(k)\mid ^2\cos ^2\varphi +\mid a_2(k)\mid
^2\sin ^2\varphi +\mid a_1(k)a_2(k)\mid \sin \left[ \alpha _1(k)-\alpha
_2(k)\right] \sin 2\varphi ,
$$
$$
\mid a_2(k+1)\mid ^2=\mid a_1(k)\mid ^2\sin ^2\varphi +\mid a_2(k)\mid
^2\cos ^2\varphi -\mid a_1(k)a_2(k)\mid \sin \left[ \alpha _1(k)-\alpha
_2(k)\right] \sin 2\varphi .\eqno{(2.8)}
$$
After the measurement in the time moment $t=k\tau $ the phase difference $
\alpha _1(k)-\alpha _2(k)$, according to the above statement, is random and
the contribution of the last term in expressions (2.8) on the average for
the large number of iterations equals zero. This results in the equation for
the probabilities
$$
\left( \matrix{p_1(k+1)\cr p_2(k+1)}\right) ={\bf M}\left(
\matrix{p_1(k)\cr
p_2(k)}\right) ,\eqno(2.9)
$$
where
$$
{\bf M}=\left(
\matrix{\cos^2\varphi & \sin^2\varphi\cr
\sin ^2\varphi & \cos ^2\varphi}\right) \eqno(2.10)
$$
is the evolution matrix for the probabilities. The full evolution from the
initial time $t=0$ until $t=T$ with the $(n-1)$ equidistant intermediate
measurement is described by the equation
$$
\left( \matrix{p_1(n)\cr p_2(n)}\right) ={\bf M}^n\left(
\matrix{p_1(0)\cr
p_2(0)}\right) .\eqno(2.11)
$$
The result of the calculation of the matrix ${\bf M}^n$ by the method of
diagonalization of the matrix ${\bf M}$ is
$$
{\bf M}^n={\frac 12}\left(
\matrix{1+\cos ^n2\varphi & 1-\cos ^n2\varphi\cr
1-\cos ^n2\varphi & 1+\cos ^n2\varphi}\right) .\eqno(2.12)
$$
From Eqs. (2.11) and (2.12) we recover the quantum Zeno effect obtained by
the density matrix technique [3-6]: if initially the system is in the state $
\Psi _1$, than the result of the evolution until the time moment $T=n\tau
=\pi /\Omega $ (after the $\pi $-pulse) with the $(n-1)$ intermediate
measurement will be characterized by the probabilities $p_1(T)$ and $p_2(T)$
for finding the system in the states $\Psi _1$ and $\Psi _2$ respectively:
$$
p_1(T)=\frac 12(1+\cos ^n2\varphi )\simeq \frac 12(1+e^{-\frac{\pi ^2}{2n}
})\simeq 1-\frac{\pi ^2}{4n}\rightarrow 1,
$$
$$
p_2(T)=\frac 12(1-\cos ^n2\varphi )\simeq \frac 12(1-e^{-\frac{\pi ^2}{2n}
})\simeq \frac{\pi ^2}{4n}\rightarrow 0,~n\rightarrow \infty .\eqno(2.13)
$$
We see that results of equations (2.11)-(2.13) represent the inhibition of
the quantum dynamics by measurements and coincide with those obtained by the
density matrix technique [3-6]. This also confirms correctness of the
proposition that the act of the measurement can be represented as
randomization of the amplitudes' phases. Further we will use this
proposition and the same method for the analysis of the repeated measurement
influence for the quantum dynamics of multilevel systems which classical
counterparts exhibit chaos. We restrict ourselves to the strongly driven by
a periodic force systems with one degree of freedom. The investigation is
also based on the mapping equations of motion for such systems.

\begin{center}
{\bf III. QUANTUM MAPS FOR MULTILEVEL SYSTEMS}
\end{center}

In general the classical equations of motion are nonintegrable and the
Schr\"odinger equation for strongly driven systems can not be solved
analytically. However, mapping forms of the classical and quantum equations
of motion greatly facilitates the investigation of stochasticity and
quantum--classical correspondence for the chaotic dynamics. From the
standpoint of an understanding of the manifestation of the measurements for
the dynamics of the multilevel systems the region of large quantum numbers
is of greatest interest. Here we can use the quasiclassical approximation
and convenient variables are the angle $\theta $ and the action $I$.
Transition from classical to the quantum (quasiclassical) description can be
undertaken replacing $I$ by the operator $\hat I=-i\frac \partial {\partial
\theta }$ [22, 23]. (We use the system of units with $\hbar =1$). One of the
simplest systems in which the dynamical chaos and its quantum localization
can be observed is a system with one degree of freedom described by the
unperturbed Hamiltonian $H_0(I)$ and driven by periodic kicks. The full
Hamiltonian $H$ of the driven system can be represented as
$$
H(I,\theta ,t)=H_0(I)+k\cos \theta \sum_j\delta \left( t-j\tau \right)
\eqno{(3.1)}
$$
where $\tau $ and $k$ are the period and strength of the perturbation,
respectively.

The intrinsic frequency of the unperturbed system is $\Omega =dH_0/dI.$ In
particular, for a linear oscillator $H_0=\Omega I$. For $H_0=I^2/2$ we have
widely investigated rotator which results to the so-called standard map [12,
24], while the Hamiltonian (3.1) with $H_0=\omega /\left[ 2\omega \left(
I_0+I\right) \right] ^{1/2}$ and $k=2\pi bF/\omega ^{5/3}$ (where $b\simeq
0.411$) models the highly excited atom in a monochromatic field of the
strength $F$ and frequency $\omega $ [23,25-27].

Integration of the classical equations of motion for the Hamiltonian (3.1)
over the perturbation period $\tau $ leads to the classical map for the
action and angle
$$
I_{j+1}=I_j+k\sin \theta ,
$$
$$
\theta _{j+1}=\theta _j+\tau \Omega \left( I_{j+1}\right) .\eqno{(3.2)}
$$
In the case of rotator the unperturbed frequency is $\Omega \left(
I_{j+1}\right) =I_{j+1}$ and the map (3.2) coincides with the investigated
in great detail standard map [12,22,24].

For the derivation of the quantum equations of motion we expand the state
function $\psi (\theta ,t)$ of the system through the quasiclassical
eigenfunctions, $\varphi _n(\theta )=e^{in\theta }/\sqrt{(}2\pi )$, of the
Hamiltonian $H_0$,
$$
\psi (\theta ,t)=(2\pi )^{-1/2}\sum\limits_na_n(t)i^{-n}e^{-in\theta }.
\eqno(3.3)
$$
Here the phase factor $i^{-n}$ is introduced for the maximal simplification
of the quantum map. Integrating the Schr\"odinger equation over the period $
\tau $, we obtain the following maps for the amplitudes before the
appropriate kicks [21, 23]
$$
a_m(t_{j+1})=e^{-iH_0(m)\tau }\sum\limits_na_n(t_j)J_{m-n}(k),~~~t_j=j\tau
\eqno(3.4)
$$
where $J_m(k)$ is the Bessel function.

The form (3.4) of the map for the quantum dynamics is rather common: similar
maps can be derived for the monochromatic perturbations as well, e.g. for an
atom in a microwave field [23, 27]. A particular case of map (3.4),
corresponding to the model of a quantum rotator $H=I^2/2$, has been
comprehensively investigated with the aim of determining the relationship
between classical and quantum chaos [12,22,24]. It has been shown that under
the onset of dynamical chaos at $K\equiv \tau k>K_c=0.9816$, motion with
respect to $I$ is not bounded and it is of a diffusion nature in the
classical case, while in the quantum description diffusion with respect to $m
$ is bounded, i.e. the diffusion ceases after some time and the state of the
system localizes exponentially. The exponential localization length $\lambda
$ of the quantum state is usually defined as follows:
$$
\lim \limits_{N\to \infty }|a_m(N\tau )|^2\sim \exp (-{\frac{2|m-m_0|}\lambda
})\eqno(3.5)
$$
where $m_0$ is the initial action. It has been shown in papers [10-12] that
for a quantum rotator the localization length is $\lambda \simeq k^2/2$ .
The effect of quantum limitation of dynamic chaos is extremely interesting
and important. It reveals itself for many quantum systems which classical
counterparts exhibit chaos. Note, that for the rotator the exact quantum
description coincides with the quasiclassical one.

Classical dynamics of the system described by map (3.2) in the case of
global distinct stochasticity is diffusion-like with the diffusion
coefficient in the $I$ space
$$
B(I)=\overline{\left( \Delta I\right) ^2}/2\tau =k^2/4\tau .\eqno(3.6)
$$

From equations (3.4) we obtain the transitions probabilities $P_{n,m}$
between $n$ and $m$ states during the period $\tau $:
$$
P_{n,m}=J_{m-n}^2(k).\eqno(3.7)
$$
Using the expression $\sum_nn^2J_n^2(k)=k^2/2$ and approximation of the
uncorrelated transitions we can formally evaluate the local quantum
diffusion coefficient in the $n$ space [21,25,26]
$$
B(n)={\frac 1{2\tau }}\sum\limits_m(m-n)^2J_{m-n}^2(k)={\frac{k^2}{4\tau }}
\eqno(3.8)
$$
Therefore, the expression for the local quantum diffusion coefficient
coincides with the classical equation (3.6).

However, it turns out that such a quantum diffusion takes place only for
some finite time $t\leq t^{*}\simeq \tau k^2/2$ [28] after which an
essential decrease of the diffusion rate is observed. Such a behavior of
quantum systems in the region of strong classical chaos is called ''the
quantum suppression of classical chaos'' [10,11]. This phenomenon turns out
to be typical for models (3.1) with nonlinear Hamiltonians $H_0(I)$ and for
other quantum systems. Thus, the diffusion coefficient (3.8) derived in the
approximation of uncorrelated transitions (3.7) does not describe the true
quantum dynamics in the energy space.

The quantum interference effect is essential for such dynamics and it
results in the quantum evolution being quantitatively different from the
classical motion. Quantum equations of motion, i.e. the Schr\"odinger
equation and the maps for the amplitudes, are linear equations with respect
to the wave function and probability amplitudes. Therefore, the quantum
interference effect manifests itself even for quantum dynamics of the
systems, the classical counterparts of which are described by nonlinear
equations and chaotic dynamics of them exhibit a dynamical chaos. On the
other hand, quantum equations of motion are very complex as well. Thus, the
Schr\"odinger equation is a partial differential equation with the
coordinate and time dependent coefficients, while the system of equations
for the amplitudes is the infinite system of equations. Moreover, for the
nonlinear Hamiltonian $H_0(m)$ the phases' increments, $H_0(m)\tau $, during
the free motion between two kicks while reduced to the basic interval $
\left[ 0,2\pi \right] $ are the pseudorandom quantities as a function of the
state's quantum number $m$. This causes a very complicated and irregular
quantum dynamics of the classically chaotic systems. We observe not only
very large and apparently irregular fluctuations of probabilities of the
states' occupation but also almost irregular fluctuations in time of the
momentum dispersion (see curves (a) in the figures 1 and 2).

However, the quantum dynamics of such driven by the external periodic force
systems is coherent and the evolution of the amplitudes $a_m(t_{j+1})$ in
time is linear: they are defined by the linear map (3.4) with the time
independent coefficients. The nonlinearity of the Hamiltonian $H_0(I)$,
being the reason of the classical chaos, causes the pseudorandom nature of
the increments of the phases, $H_0(m)\tau $, as a function of the state's
number $m$ (but constant in time). These increments of the phases remain the
same for each kick. So, the dynamics of the amplitudes $a_m(t_{j+1})=\left|
a_m(t_{j+1})\right| e^{i\alpha _m\left( t_{j+1}\right) }$ and of their
phases, $\alpha _m\left( t_{j+1}\right) $, is strongly deterministic and
non-chaotic but very complicated and apparently irregular. For instance, the
phases $\alpha _m\left( t_{j+1}\right) $ are phases of the complex
amplitudes, $a_m(t_{j+1})$, which are linear combinations (3.4) of the
complex amplitudes, $a_n(t_j)$, before the preceding kick with the
pseudorandom coefficients, $e^{-iH_0(m)\tau }J_{m-n}(k)$. Nevertheless, the
iterative equation (3.4) is a {\it linear transformation with coefficients
regular in time.} That is why, we observe for such dynamics the
quasiperiodic reversible in the time evolution [12] with the quantum
localization of the pseudochaotic motion.

In paper [23] it has been demonstrated that this peculiarity of the
pseudochaotic quantum dynamics is indeed due to the pseudorandom nature of
the phases, $H_0(m)\tau $, in Eq. (3.4) as a function of the eigenstate's
quantum number $m$ (but not of the kick's number $j$). Replacing the
multipliers $\exp \left[ -iH_0(m)\tau \right] $ in Eq. (3.4) by the
expressions $\exp \left[ -i2\pi g_m\right] $, where $g_m$ is a sequence of
random numbers that are uniformly distributed in the interval $\left[
0,1\right] $, we observe the quantum localization as well [23]. The
essential point here is the independence of the phases $H_0(m)\tau $ or $
2\pi g_m$ on the step of iteration $j$ or time $t$. This is the main core of
difference from the randomness of the phases due to the measurements under
consideration in the next Section.

\begin{center}
{\bf IV. INFLUENCE\ OF\ REPETITIVE\ MEASUREMENT\ ON\ THE\ QUANTUM\ DYNAMICS}
\end{center}

Each measurement of the system's state projects the state into one of the
energy state $\varphi _m$ with the definite $m$. Therefore, if we make a
measurement of the system after the kick $j$ but before the next $(j+1)$
kick we will find the system in the states $\varphi _m$ with the appropriate
probabilities $p_m(j)=\left| a_m(t_j)\right| ^2$.

In principle, such a measurement can be performed in the same way as in the
experiment of Itano {\it et al.} [4], i.e. by the short-impulse laser
excitation of the system from state $\varphi _m$ to some higher state
followed by the irreversible return of the system to the same state $\varphi
_m$ with registration of the state's population by photon counting. After
the measurement of the state's $\varphi _m$ population, the probability of
finding the system in the state $\varphi _m$ coincides with that before the
measurement. However there is no interference between the state's $\varphi _m
$ amplitude $\tilde a_m(t_j)$ after the measurement and amplitudes of other
states, $a_n(t_j)$, i.e. the cross terms containing the amplitude $\tilde a
_m(t_j)$ in the expressions for probabilities vanish. But interference
between the unmeasured states remains and the cross terms containing the
amplitudes of the unmeasured states do not vanish.

In the calculations of the system's dynamics the influence of the
measurements can be taken into account in the same way as in the Section II,
i.e. through randomization of phases of the amplitudes after the measurement
of the appropriate state's populations. The phases of amplitudes after the
measurement are completely random and uncorrelated with the phases before
the measurement, after another measurements and with the phases of other
measured or unmeasured states. Therefore, after the full measurement of the
system after the kick $j$, all phases of the amplitudes $a_m(t_j)$ are
random. So, this full measurement of the system's state influences on the
further dynamics of the system through the randomization of the phases of
amplitudes (see Section II for analogy). This fact can be expressed by
replacement in Eqs. (3.4) of the amplitudes $a_m(t_{j+1})$ by the amplitudes
$e^{i\beta _m(t_{j+1})}a_m(t_{j+1})$ with the random phases $\beta
_m(t_{j+1})$. The essential point here is that the phases $\beta _m(t_{j+1})$
are different, uncorrelated for the different measurements, i.e. for
different time moments of the measurement $t_{j+1}$. This is the principal
difference of the random phases $\beta _m(t_{j+1})$ from the phases $
H_0(m)\tau $ in Eqs. (3.4) which are pseudorandom variables as functions of
the eigenstate's quantum number $m$ (but not of the time moment $t_{j+1}$).

In such a way, introducing the appropriate random phases we can analyze the
influence on the system's dynamics of the full measurements of the system's
state performed after every kick, after every $N$ kicks or of the
measurements of the population probabilities just of some states, e.g. only
of the initial state. Note that there is no need to measure more frequently
than after every kick because the results of the subsequent measurements
before the next kick repeat the results of the preceding measurements (after
the last kick).

Instead of representing the detailed quantum dynamics expressed as the
evolution of all amplitudes in the expansion of the wave function (3.3) we
can represent only dynamics of the momentum dispersion
$$
\left\langle (m_j-m_0)^2\right\rangle =\sum\limits_m\left( m-m_0\right)
^2\left| a_m\left( t_j\right) \right| ^2\eqno{(4.1)}
$$
where $m_0$ is the initial momentum (quantum number). Such a representation
of the dynamics is simpler, more picturesque and more comfortable for
comparison with the classical dynamics.

In figures 1 and 2 we show the results of numerical analysis of the
influence of measurements of the system's state on the quantum dynamics of
the rotator and of the system with random distribution of energy levels,
i.e. for random phases $H_0(m)\tau $ in Eqs. (3.4) as a function of the
eigenstate's quantum number $m$. We see that quantum diffusion-like dynamics
of the systems without measurements, represented by curves (a), after
sufficiently short time $t^{*}\simeq \tau k^2/2$ (of the order of $50\tau $
in our case) ceases and the monotonic increase of the momentum dispersion $
\left\langle (m-m_0)^2\right\rangle \simeq 2Bt={\frac{k^2}{2\tau }t}$ for
time $t\ll t^{*}$ turns for the time $t\gg t^{*}$ into the stationer (on the
average for same time interval $\Delta t\geq t^{*}$) distribution with the
momentum dispersion $\left\langle (m_{st}-m_0)^2\right\rangle \simeq \lambda
^2/2\simeq k^4/8$. This is a demonstration of the effect of quantum
suppression of the classical chaos.

In the case of measurement of the only initial, $\varphi _{500}$, state's
population after every kick (which technically is achieved by introduction
of the random phase $\beta _{500}\left( t_{j+1}\right) $ after every kick $j$
) we observe monotonic, though slow, increase of the momentum dispersion for
very long time, until $t\sim 600\tau $ in our case (curves (b) in figures 1
and 2). After such time the population of the initial state on the average
becomes very small and measurements of this state's population almost does
not influence on the systems dynamics.

The dynamics with measurements of all states every $200$ kicks represented
by curves (c) is a staircase-like: fast increase of the momentum dispersion
after the immediate measurement turns into the quantum suppression of the
diffusion-like motion for $\Delta t\geq t^{*}$ until the next measurement
destroys the quantum interference and induces the succeeding diffusion-like
motion.

The quantum dynamics of the kicked rotator or some other system with
measurements of all states' populations after every kick as represented by
the curves (d) is essentially classical-like: the momentum dispersion
increases linearly in time with the classical diffusion coefficient (3.6)
for all time of the calculation.

Theoretically such differences of dynamics can be understood from the
iterative equations for the momentum dispersion:
$$
\left\langle (m_{j+1}-m_0)^2\right\rangle =\sum\limits_m\left( m-m_0\right)
^2\left| a_m\left( t_{j+1}\right) \right| ^2,\eqno{(4.2)}
$$
where
$$
\left| a_m\left( t_{j+1}\right) \right| ^2=\sum_{n,n^{^{\prime
}}}J_{m-n}\left( k\right) J_{m-n^{^{\prime }}}\left( k\right) a_n\left(
t_j\right) a_{n^{^{\prime }}}^{*}\left( t_j\right) .\eqno{(4.3)}
$$
Substitution of Eq. (4.3) into Eq. (4.2) yields
$$
\left\langle (m_{j+1}-m_0)^2\right\rangle =\sum\limits_{m,n}\left(
m-m_0\right) ^2J_{m-n}^2\left( k\right) \left| a_n\left( t_j\right) \right|
^2+2\sum\limits_{m,n}\sum\limits_{n^{^{\prime }}<n}\left( m-m_0\right)
^2J_{m-n}\left( k\right) J_{m-n^{^{\prime }}}\left( k\right) Re\left[
a_n\left( t_j\right) a_{n^{^{\prime }}}^{*}\left( t_j\right) \right] .
\eqno{(4.4)}
$$

For the random phase differences of the amplitudes $a_n\left( t_j\right) $
and $a_{n^{^{\prime }}}^{*}\left( t_j\right) $ with $n^{^{\prime }}\neq n$,
which is a case after the measurement of the system's state, the second term
of Eq. (4.4) on the average equals zero (see Section II for clarification).
Then from Eq. (4.4) we have
$$
\left\langle (m_{j+1}-m_0)^2\right\rangle =\sum\limits_n\left| a_n\left(
t_j\right) \right| ^2\sum\limits_m\left( m-m_0\right) ^2J_{m-n}^2\left(
k\right) =\sum\limits_m\left| a_m\left( t_j\right) \right| ^2\left(
m^2-m_0^2+\frac{k^2}2\right) =\left\langle (m_j-m_0)^2\right\rangle +\frac{
k^2}2.\eqno{(4.5)}
$$
In the derivation of Eq. (4.5) we have used the summation expressions
$$
\sum\limits_mmJ_{m-n}^2\left( k\right) =0\quad \mbox{and}\quad
\sum\limits_mm^2J_{m-n}^2\left( k\right) =n^2+\frac{k^2}2.
$$

Therefore, according to Eq. (4.5) for the uncorrelated phases of the
amplitudes $a_n\left( t_j\right) $ and $a_{n^{^{\prime }}}^{*}\left(
t_j\right) $ with $n^{^{\prime }}\neq n$ the dispersion of the momentum as a
result of every kick increases on the average in the magnitude $k^2/2$, the
same as for the classical dynamics. For dynamics of isolated quantum systems
without measurements or unpredictable interaction with the environment the
second term of Eq. (4.4), due to the quantum interference between the
amplitudes of different states arisen from the same initial states'
superposition, compensate (on the average for sufficiently large time
interval $\Delta t\geq t^{*}$) the first term of Eq.(4.4) and so the quantum
suppression of dynamics may be observed.

Similar analysis can be used for the investigation of the influence of the
measurements on quantum dynamics of another quantum systems with quantum
localization of the classical chaos as well.

As it has already been stated above the influence of the repetitive
measurement on quantum dynamics is closely related with the affect of the
unpredictable interaction between the system and the environment. It should
be noticed that in general for the analysis of the measurement effect and to
facilitate the comparison between quantum and classical dynamics of chaotic
systems it is convenient to employ the Wigner representation, $\rho _W\left(
x,p,t\right) $, of the density matrix [19, 29]. The Wigner function of the
system with the Hamiltonian of the form $H=p^2/2m+V\left( x,t\right) $
evolves according to equation
$$
\frac{\partial \rho _W}{\partial t}=\left\{ H,\rho _W\right\} _M\equiv
\left\{ H,\rho _W\right\} +\sum\limits_{n\geq 1}\frac{\hbar ^{2n}\left(
-1\right) ^n}{2^{2n}\left( 2n+1\right) !}\frac{\partial ^{2n+1}V}{\partial
x^{2n+1}}\frac{\partial ^{2n+1}\rho _W}{\partial p^{2n+1}},\eqno{(4.6)}
$$
where by $\left\{ ...\right\} _M$ and $\left\{ ...\right\} $ are denoted the
Moyal and the Poisson brackets, respectively. The terms in Eq. (4.6)
containing Planck's constant and higher derivatives give the quantum
corrections to the classical dynamics generated by the Poisson brackets. In
the region of regular dynamics one can neglect the quantum corrections for
very long time if the characteristic actions of the system are large. For
classically chaotic motion the exponential instabilities lead to the
development of the fine structure of the Wigner function and exponential
growth of its derivatives. As a result, the quantum corrections become
significant after relatively short time even for macroscopic bodies [19,
28]. The extremely small additional diffusion-like terms in Eq. (4.6), which
reproduce the effect of interaction with the environment or frequent
measurement, prohibits development of the Wigner function's fine structure
and removes barriers posed by classical chaos for the correspondence
principle [19, 29].

\begin{center}
{\bf V. CONCLUSIONS}
\end{center}

From the above analysis we can conclude that the influence of the repetitive
measurement on the dynamics of the quasiclassical multilevel systems with
the quantum suppression of the classical chaos is opposite to that for the
few-level quantum system. The repetitive measurement of the multilevel
systems results in delocalization of the states superposition and
acceleration of the chaotic dynamics. In the limit of the frequent full
measurement of the system's state the quantum dynamics of such systems
approaches the classical motion which is opposite to the quantum Zeno
effect: inhibition or even prevention of time evolution of the system from
an eigenstate of observable into a superposition of eigenstates by repeated
frequent measurement. Therefore, we can call this phenomenon the 'quantum
anti-Zeno effect'.

It should be noted that the same effect can be derived without the {\it ad
hoc} collapse hypothesis but from the quantum theory of irreversible
processes, in analogy with the method used in the papers [6, 9]. Even the
simplest detector follows irreversible dynamics due to the coupling to the
multitude of vacuum modes which results in the randomization of the quantum
amplitudes' phases, decay of the off-diagonal matrix elements of the density
matrix and to smoothing of the fine structure of the Wigner distribution
function, i.e. just what we need to obtain the classical equations of motion.

So, the quantum-classical correspondence problem caused by the chaotic
dynamics is closely related with the old problem of measurement in quantum
mechanics. In the case of frequent measurement or unpredictable interaction
with the environment the quantum dynamics of the quasiclassical systems
approaches the classical-like motion.

\begin{center}
{\bf ACKNOWLEDGMENT }
\end{center}

The research described in this publication was made possible in part by
support of the Alexander von Humboldt Foundation.

\vskip\baselineskip

\begin{center}
---------------------------------------
\end{center}

\begin{enumerate}
\item  L. A. Khalfin, Zh. Eksp. Teor. Fiz. {\bf 33}, 1371 (1958) [Sov. Phys.
JETP {\bf 6}, 1503 (1958)]; J. Swinger, Ann. Phys. {\bf 9}, 169 (1960); L.
Fonda, G. C. Ghirardi, and A. Rimini, Rep. Prog. Phys. {\bf 41}, 587 (1978);
G.-C. Cho, H. Kasari, and Y. Yamaguchi, Prog. Theor. Phys. {\bf 90}, 803
(1993).

\item  B. Misra and E. C. G. Sudarshan, J. Math. Phys. {\bf 18}, 756 (1976);
C. B. Chiu, E. C. G. Sudarshan, and B. Misra, Phys. Rev. D {\bf 16}, 520
(1977).

\item  R. J. Cook, Phys. Scr. {\bf T21}, 49 (1988).

\item  W. M. Itano, D. J. Heinzen, J. J. Bollinger, and D. J. Wineland,
Phys. Rev. A {\bf 41}, 2295 (1990).

\item  E. Joos, Phys. Rev. D {\bf 29}, 1626 (1984); T. Petrosky, S. Tasaki,
and I. Prigogine, Phys. Lett. A {\bf 151}, 109 (1990); P. Knight, Nature
{\bf 344}, 493 (1990); A. Schenzle, Contemp. Phys. {\bf 37} 303 (1996).

\item  V. Frerichs and A. Schenzle, Phys. Rev. A {\bf 44}, 1962 (1991).

\item  S. Pascazio and M. Namiki, Phys. Rev. A {\bf 50}, 4582 (1994); H.
Nakazato, M. Namiki, S. Pascazio, and H. Rauch, Phys. Lett. {\bf A217}, 203
(1996).

\item  Y. Aharonov and M. Vardi, Phys. Rev. D {\bf 21}, 2235 (1980).

\item  T. P. Altenm\"uller and A. Schenzle, Phys. Rev. A {\bf 48}, 70 (1993).

\item  G. Casati, B. V. Chirikov, J. Ford, and F. M. Izrailev, in {\it
Stochastic Behavior in Classical and Quantum Hamiltonian Systems,} edited by
G. Casati and J. Ford Lecture Notes in Physics Vol. 93 (Springer-Verlag.
Berlin, 1979), p. 334.

\item  G. Casati, B. V. Chirikov, D. L. Shepelyansky, and I. Guarneri, Phys.
Rep. {\bf 154}, 77 (1987).

\item  F. M. Izrailev, Phys. Rep. {\bf 196}, 299 (1990).

\item  E. Ott, T. M. Antonsen, Jr., and J. D. Hanson, Phys. Rev. Lett. {\bf
53}, 2187 (1984).

\item  S. Adachi, M. Toda, and K. Ikeda, J. Phys. A: Math. Gen. {\bf 22},
3291 (1989).

\item  T. Dittrich and R. Graham, Ann. Phys. {\bf 200}, 363 (1990).

\item  R. Bl\"umel, A. Buchleiter, R. Graham, L. Sirko, U. Smilansky, and H.
Walther, Phys. Rev. A {\bf 44}, 4521 (1991).

\item  P. Goetsch and R. Graham, Phys. Rev. E {\bf 50}, 5242 (1994).

\item  D. Cohen and S. Fishman, Phys. Rev. Lett. {\bf 67}, 1945 (1991).

\item  W. H. Zurek, Phys. Today {\bf 44}, 36 (1991); W. H. Zurek and J. P.
Paz, Phys. Rev. Lett. {\bf 72}, 2508 (1994).

\item  K. Shiokawa and B. L. Hu, Phys. Rev. E {\bf 52}, 2497 (1995).

\item  B. Kaulakys, in {\it Quantum Communications and Measurement}, eds V.
P. Belavkin, O. Hirota, and R. L. Hudson (Plenum Press, 1995), p. 193,
quant-ph/9503018; V. Gontis and B. Kaulakys, {\it J. Tech. Phys.} {\bf 38},
223 (1997).

\item  G. M. Zaslavskii, {\it Stochastic Behavior of Dynamical Systems}
(Nauka, Moscow, 1984; Harwood, New York, 1985).

\item  V. G. Gontis and B. P. Kaulakys, Liet. Fiz. Rink. {\bf 28}, 671
(1988) [Sov. Phys.-Collec. {\bf 28}(6), 1 (1988)].

\item  A. J. Lichtenberg and M. A. Lieberman, {\it Regular and Stochastic
Motion} ( Springer-Verlag, New York, 1983).

\item  V. Gontis and B. Kaulakys, J. Phys. B: At. Mol. Opt. Phys. {\bf 20},
5051 (1987).

\item  B. Kaulakys, V. Gontis, G. Hermann, and A. Scharmann, Phys. Lett. A
{\bf 159}, 261 (1991).

\item  V. Gontis and B. Kaulakys, Lithuanian J. Phys. {\bf 31}, 75 (1991).

\item  G. Casati and B. Chirikov, ''The legacy of chaos in quantum
mechanics'', in {\it Quantum chaos: between order and disorder}, Ed. G.
Casati and B. V. Chirikov (Cambridge University, 1994), p.3.

\item  B. Kaulakys, Lithuanian J. Phys. {\bf 36}, 343 (1996); B. Kaulakys,
quant-ph/9610041.
\end{enumerate}

\newpage

Caption for figure to the paper\vspace{0.5cm}

\begin{center}
{\Large B. Kaulakys and V. Gontis }

{\Large 'Quantum anti-Zeno effect'}\vspace{0.5cm}
\end{center}

Fig. 1 Dependence of the dimensionless momentum dispersion, $\left\langle
(m-m_0)^2\right\rangle $, as defined by Eq. (4.2) for the quantum rotator
with $m_0=500$, $\tau =1$ and $k=10$ on the discrete dimensionless time $j$
for the dynamics according to Eq. (3.4): (a) without the intermediate
measurements, (b) with measurements of the initial state, $\varphi _{500}$,
after every kick, (c) with measurements of all states every $200$ kicks and
(d) with measurements of all states after every kick.\vskip 2
\normalbaselineskip

Fig. 2. Same as in Fig. 1 but for the system with random distribution of
energy levels, i.e. for random phases $H_0(m)\tau $ in Eqs. (3.4) defined as
$2\pi g_m$ where $g_m$ is a sequence of random numbers that are uniformly
distributed in the interval $\left[ 0,1\right] $.

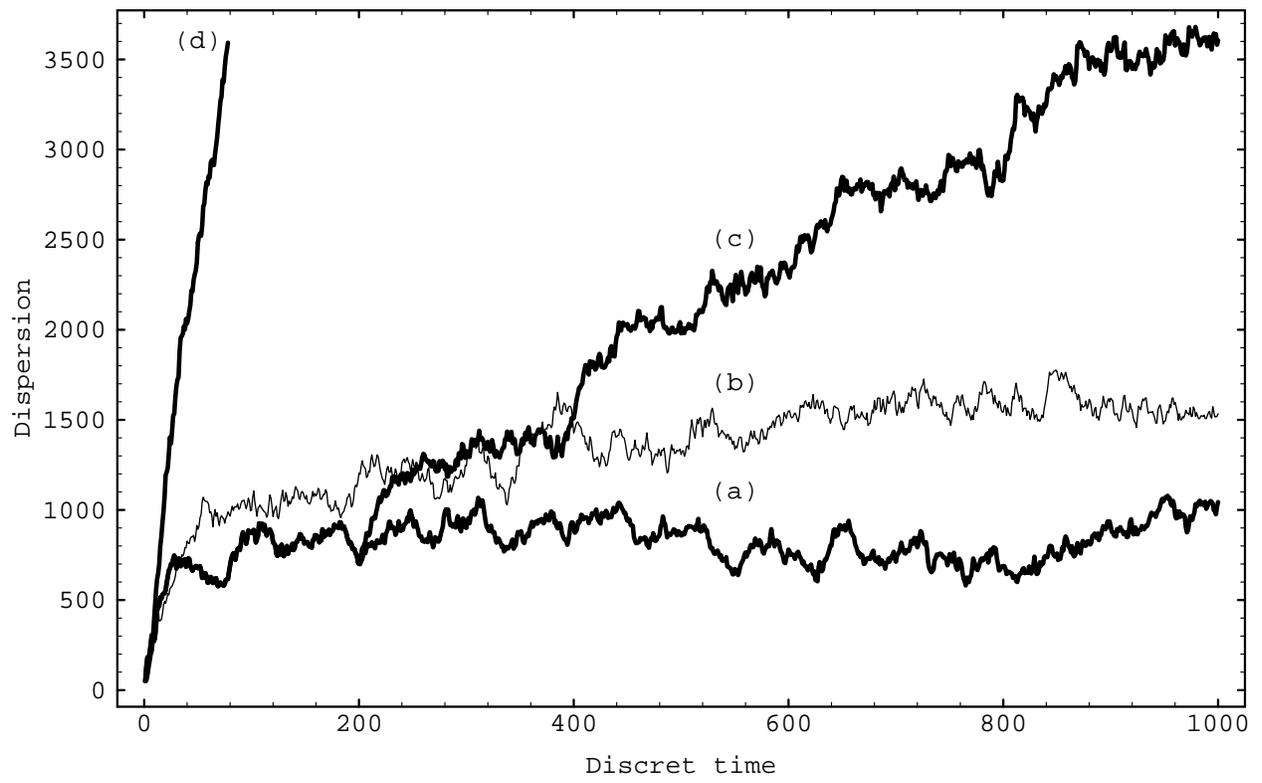
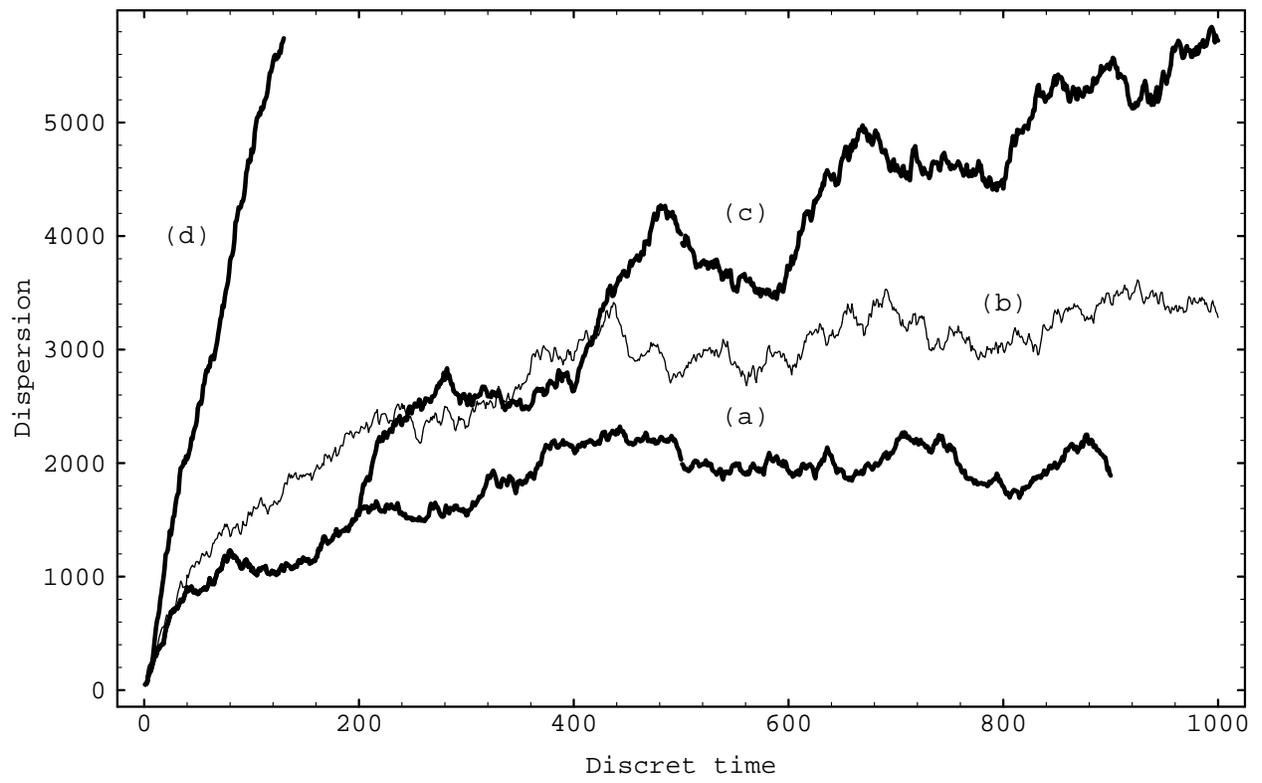
\end{document}